\documentclass{ws-p8-50x6-00.new}

\begin{document}

\newcommand{\beq}{\begin{equation}}
\newcommand{\eeq}{\end{equation}}
\newcommand{\barr}{\begin{eqnarray}}
\newcommand{\earr}{\end{eqnarray}}

\newcommand{\andy}[1]{ }
\def\cH{{\mathcal{H}}}
\def\cV{{\mathcal{V}}}
\def\cU{{\mathcal{U}}}
\def\cP{{\mathcal{P}}}
\def\cN{{\mathcal{N}}}
\def\cS{{\mathcal{S}}}
\def\bra#1{\langle #1 |}
\def\ket#1{| #1 \rangle}
\def\As{{\mathcal{A}}}
\def\Ord{\mathrm{O}}

\title{Quantum Zeno effect, adiabaticity\\ and dynamical superselection rules}

\author{Paolo Facchi}

\address{Dipartimento di Fisica, Universit\`a di Bari  I-70126 Bari, Italy
\\
and Istituto Nazionale di Fisica Nucleare, Sezione di Bari,
I-70126 Bari, Italy }


\maketitle

\abstracts{ The evolution of a quantum system undergoing very
frequent measurements takes place in a proper subspace of the
total Hilbert space (quantum Zeno effect). When the measuring
apparatus is included in the quantum description, the Zeno effect
becomes a pure consequence of the dynamics. We show that for
continuous measurement processes the quantum Zeno evolution
derives from an adiabatic theorem. The system is forced to evolve
in a set of orthogonal subspaces of the total Hilbert space and a
dynamical superselection rule arises. The dynamical properties of
this evolution are investigated and several examples are
considered. }

%\pacs{PACS numbers: 03.65.Xp}

\section{Introduction }
 \label{sec-introd}
 \andy{intro}

The quantum Zeno effect\cite{Beskow,Misra} is a direct consequence
of general features of the Schr\"odinger equation that yield
quadratic behavior of the survival probability at short
times.\cite{strev,zenoreview} It consists in the hindrance of the
evolution of a quantum system when very frequent measurements are
performed, in order to ascertain whether it is still in its
initial state.

Following an interesting idea by Cook,\cite{Cook} both the
experimental and theoretical investigations of the last decade
have dealt with oscillating (mainly, two-level)
systems.\cite{Itano} However, a few years ago, the presence of a
short-time quadratic region was experimentally confirmed for an
unstable quantum mechanical system\cite{Wilkinson} and the
existence of the Zeno effect (as well as its
inverse\cite{antiZeno,heraclitus}) has been recently
proved.\cite{raizenlatest}

Interestingly, the quantum Zeno effect (QZE) does not necessarily
freeze everything. On the contrary, for frequent projections onto
a multi-dimensional subspace, the system can evolve away from its
initial state, although it remains in the subspace defined by the
``measurement." This continuing time evolution within the
projected subspace has been recently investigated and called {\em
quantum Zeno dynamics}.\cite{compactregularize} It involves
interesting, yet unsettled, physical and
mathematical\cite{Friedman72,Gustafson,Misra} issues.

The above-mentioned investigations deal with ``pulsed"
measurements, according to von Neumann's projection
postulate.\cite{von} However, from a physical point of view, a
``measurement" is nothing but an interaction with an external
system (another quantum object, or a field, or simply a different
degree of freedom of the very system investigated), playing the
role of apparatus. In this sense von Neumann's postulate can be
considered as a useful shorthand notation, summarizing the effect
of the quantum measurement.

By including the apparatus in the quantum description, several
authors,\cite{Itano,Peres80,PeresKraus} during the last two
decades, have demonstrated the QZE without making use of
projection operators (and non-unitary dynamics). In particular,
the QZE has been reformulated in terms of ``continuous"
measurements,\cite{zenoreview,Napoli} obtaining the same physical
effects (as well as a quantitative comparison with the ``pulsed"
situation\cite{Schulman98}) in terms of a continuous (eventually
strong) coupling to an external agent.

The studies of the last few years pave the way to interesting
possible applications of the QZE. Indeed, we have a physical and
mathematical framework that enable us to analyze the modification
of the evolution of a quantum system and possibly to tailor the
interaction in order to slow the evolution down (or eventually
accelerate it). The potential importance of such a scheme cannot
be underestimated.

It is therefore important to understand in more details which
features of the coupling between the ``observed" system and the
``measuring" apparatus are needed to obtain a QZE. In other words,
one wants to know when an external quantum system can be
considered a good apparatus and why. The purpose of the present
article is to clarify these issues and cast the quantum Zeno
evolution in terms of an adiabatic theorem. We will show that the
evolution of a quantum system under the action of a continuous
measurement process can be profoundly modified: the system is
forced to evolve in a set of orthogonal subspaces of the total
Hilbert space and an effective superselection rule arises in the
strong coupling limit. These \textit{quantum Zeno
subspaces}\cite{theorem} are just the eigenspaces (belonging to
different eigenvalues) of the Hamiltonian describing the
interaction between the system and the apparatus: they are
subspaces that the measurement process is able to distinguish.

The general ideas will be applied to some relevant examples. Some
interesting issues and possible applications will be discussed in
details.

\section{Pulsed measurements}
\label{sec-pulsed}
\andy{pulsed}

Let Q be a quantum system, whose states belong to the Hilbert
space ${\cal H}$ and whose evolution is described by the unitary
operator $U(t)=\exp(-iHt)$, where $H$ is a time-independent
lower-bounded Hamiltonian. Let $P$ be a projection operator and
$\textrm{Ran}P=\cH_P$ its range. We assume that the initial
density matrix $\rho_0$ of system Q belongs to ${\cal H}_P$:
\andy{inprep}
\beq
\rho_0 = P \rho_0 P , \qquad \mathrm{Tr} [ \rho_0 P ] = 1 .
\label{eq:inprep}
\eeq
Under the action of the Hamiltonian $H$ (i.e., if no measurements
are performed in order to get information about the quantum
state), the state at time $t$ reads
\andy{noproie}
\beq
\rho (t) = U(t) \rho_0 U^\dagger (t)
  \label{eq:noproie}
\eeq
and the \textit{survival probability}, namely the probability that
the system is still in ${\cal H}_P$ at time $t$, is
\andy{stillun}
\beq
p(t) = \mathrm{Tr} \left[ U(t) \rho_0 U^\dagger(t) P \right] .
\label{eq:stillun}
\eeq
No distinction is made between one- and many-dimensional
projections.

The above evolution is ``undisturbed," in the sense that the
quantum systems evolves only under the action of its Hamiltonian
for a time $t$, without undergoing any measurement process.
Assume, on the other hand, that we do perform a selective
measurement at time $\tau$, in order to check whether Q has
survived inside ${\cal H}_P$. By this, we mean that we select the
survived component and stop the other one. The state of Q changes
(up to a normalization constant) into
\andy{proie}
\beq
\rho_0 \rightarrow \rho(\tau) = P U(\tau) \rho_0 U^\dagger(\tau) P
\label{eq:proie}
\eeq
and the survival probability in ${\cal H}_P$ is
\andy{probini}
\barr
p(\tau) = \mathrm{Tr} \left[ U(\tau) \rho_0 U^\dagger(\tau) P
\right] = \mathrm{Tr} \left[V(\tau) \rho_0 V^\dagger(\tau)
\right],
  \qquad V(\tau) \equiv P U(\tau)P.
\label{eq:probini}
\earr
We stress that the measurement occurs instantaneously (this is the
essence of von Neumann's projection postulate).\cite{von}

The QZE is the following. We prepare Q in the initial state
$\rho_0$ at time 0 and perform a series of $P$-observations at
time intervals $\tau=t/N$. The state of Q at time $t$ reads
\andy{Nproie}
\beq
\rho^{(N)}(t) = V_N(t) \rho_0 V_N^\dagger(t) , \qquad
    V_N(t) \equiv [ P U(t/N) P ]^N
\label{eq:Nproie}
\eeq
and the survival probability in ${\cal H}_P$ is given by
\andy{probNob}
\beq
p^{(N)}(t) = \mathrm{Tr} \left[ V_N(t) \rho_0 V_N^\dagger(t)
\right].
\label{eq:probNob}
\eeq
Equations (\ref{eq:Nproie})-(\ref{eq:probNob}) are the formal
statement of the QZE, according to which very frequent
observations modify the dynamics of the quantum system: under
general conditions, if $N$ is sufficiently large, all transitions
outside ${\cal H}_P$ are inhibited. Notice that the dynamics
(\ref{eq:Nproie})-(\ref{eq:probNob}) is \textit{not reversible}.

We emphasize that close scrutiny of the features of the survival
probability has clarified that if $N$ is not too large the system
can display an inverse Zeno
effect,\cite{antiZeno,heraclitus,zenowaseda} by which decay is
accelerated. Both effects have recently been seen in the same
experimental setup.\cite{raizenlatest} We will not elaborate on
this here.

\subsection{Misra and Sudarshan's theorem }
\label{sec-MisSud}
\andy{MisSud}

In order to consider the $N \rightarrow \infty$ limit
(``continuous observation"), one needs some mathematical
requirements: assume that the limit operator
\andy{slim}
\beq
\cV (t) \equiv \lim_{N \rightarrow \infty} V_N(t)
  \label{eq:slim}
\eeq
exists (in the strong sense) for $t>0$. The final state of Q is
then
\andy{infproie}
\beq
\rho (t) = \lim_{N\to\infty} \rho^{(N)}(t)=\cV(t) \rho_0
\cV^\dagger (t)
  \label{eq:infproie}
\eeq
and the probability to find the system in $\cH_P$ is
\andy{probinfob}
\beq
\cP (t) \equiv \lim_{N \rightarrow \infty} p^{(N)}(t)
   = \mathrm{Tr} \left[ \cV(t) \rho_0 \cV^\dagger(t) \right].
\label{eq:probinfob}
\eeq
By assuming the strong continuity of $\cV(t)$ at $t=0$
\andy{phgr}
\beq
\lim_{t \rightarrow 0^+} \cV(t) = P,
\label{eq:phgr}
\eeq
one can prove that under general conditions the operators
\andy{semigr}
\beq
\cV(t) \quad \textrm{exist for all real $t$ and form a semigroup.}
\label{eq:semigr}
\eeq
Moreover, by time-reversal invariance
\andy{VVdag}
\beq
\cV^\dagger (t) = \cV(-t),
\label{eq:VVdag}
\eeq
so that $\cV^\dagger (t) \cV(t) =P$. This implies, by
(\ref{eq:inprep}), that
\andy{probinfu1}
\beq
\cP(t)=\mathrm{Tr}\left[\rho_0 \cV^\dagger(t)\cV(t)\right] =
\mathrm{Tr} \left[ \rho_0 P \right] = 1 .
\label{eq:probinfu1}
\eeq
If the particle is ``continuously" observed, in order to check
whether it has survived inside $\cH_P$, it will never make a
transition to  $\cH_P^\perp$ (QZE).

Two important remarks are now in order: first, it is not clear
whether the dynamics in the $N \to \infty$ limit is time
reversible. Although one ends up, in general, with a semigroup,
there are concrete elements of reversibility in the above
equations.\cite{compactregularize} Second, the theorem just
summarized
\textit{does not} state that the system \textit{remains} in its
initial state, after the series of very frequent measurements.
Rather, the system is left in the subspace $\cH_P$, instead of
evolving ``naturally" in the total Hilbert space $\cH$.

\subsection{Complete measurements}
 \label{sec-selmeas}
 \andy{selmeas}
Let us first consider the particular case of complete selective
measurements. $\cH_P$ has dimension 1 and the initial state is a
pure (normalized) state $\ket{a}$:
\beq
P_a=\ket{a}\bra{a},\qquad \rho_0=\ket{a}\bra{a} .
\eeq
The time evolution operator (\ref{eq:Nproie}) after $N$
measurements in a time interval $t$ reads
\beq
\label{eq:VNAs}
V_N(t) = [ P_a U(t/N) P_a]^N=[P_a \bra{a} U(t/N)\ket{a}]^N=P_a
\As(t/N)^N ,
\eeq
where $\As(t)$ is the (undisturbed) \textit{survival amplitude} in
state $\ket{a}$ at time $t$
\beq
\As(t)=\bra{a} e^{-i H t}\ket{a}.
\eeq
Therefore, in this case the problem of the existence of the limit
operator $\cV(t)$ is reduced to the existence of the limit
function $\lim_N \As(t/N)^N$. Let $\tau=t/N$ be the time interval
between two successive measurements. We can write
\barr
\As\left(\frac{t}{N}\right)^N&=&\As(\tau)^N =\exp\left[ N
\log\As(\tau)\right]=\exp\left[t\;\frac{\log\As(\tau)}{\tau}\right]
\nonumber\\
& = & \exp\left[-t
\left(\frac{\gamma(\tau)}{2}+i\omega(\tau)\right)\right],
\label{eq:Asexp}
\earr
where
\beq
\gamma(\tau)=-\frac{1}{\tau}\log|\As(\tau)|^2, \qquad
\omega(\tau)=-\frac{1}{\tau}\arg \As(\tau),
\label{eq:effective}
\eeq
and the ``observed" survival probability has a purely exponential
decay with an effective rate $\gamma(\tau)$:
\beq
p_a^{(N)}(t)=\left|\As(\tau)\right|^{2N}=\exp[- \gamma(\tau) t] .
\eeq
From (\ref{eq:VNAs}) and (\ref{eq:Asexp}) one sees that $\cV(t)$
in (\ref{eq:slim}) exists (in the strong sense), for $t>0$, if and
only if $\gamma(0^+)$ and $\omega(0^+)$ exist and are finite [or
if $\gamma(0^+)=+\infty$, the existence of the limit $\omega(0^+)$
being in this case irrelevant], and it reads
\beq
\label{eq:cVa}
\cV(t)= P_a \exp\left[-
\left(\frac{\gamma(0^+)}{2}+i\omega(0^+)\right) t\right].
\eeq
Notice that, for one-dimensional projections, when $\cV(t)$ exists
[and $\gamma(0^+)<\infty$ so that $\cV(t)\neq 0$], the strong
continuity in the origin $\cV(0^+)=P_a$ [see Eq.\ (\ref{eq:phgr})]
follows from the very existence of (\ref{eq:cVa}) and need not be
assumed as an independent hypothesis.

A sufficient condition for the existence of $\cV(t)$ is that the
initial state belongs to the domain of the Hamiltonian,
$\ket{a}\in D(H)$. Indeed, in such a case, the first and second
moment of the Hamiltonian exist,
\beq
\bra{a} H^2 \ket{a}=\| H\ket{a} \|^2 < \infty, \qquad \bra{a} H
\ket{a}=E_a < \infty,
\eeq
and the survival amplitude has the following asymptotic behavior
at short times:\footnote{More precisely, without further
information on the third moment, $\As(\tau)= 1-i E_a \tau
+\Ord(\tau^2)$, whence $\gamma(\tau)=\Ord(\tau)$.}
\beq
\label{eq:quadAs}
\As(\tau)\sim 1-i E_a \tau - \bra{a} H^2 \ket{a} \frac{\tau^2}{2},
\qquad \tau\to 0 ,
\eeq
i.e.\ the survival probability exhibits a short-time quadratic
behavior,
\beq
\label{eq:quadratic}
p_a(\tau)=|\As(\tau)|^2\sim 1-\frac{\tau^2}{\tau_{\mathrm{Z}}^2},
\qquad \tau\to 0 ,
\eeq
where
\beq
\tau_{\mathrm{Z}} \equiv \frac{1}{\sqrt{\langle a|H^2|a\rangle -
\langle a|H|a\rangle^2}}
\eeq
is called the \textit{Zeno time}. Therefore, by plugging
(\ref{eq:quadAs}) into (\ref{eq:effective}),
\beq
\gamma(\tau)\sim \frac{\tau}{\tau_{\mathrm{Z}}^2}, \qquad
\omega(\tau)\sim E_a, \qquad\quad \tau\to0,
\eeq
and one gets
\beq
\label{eq:limita}
\cV(t)=P_a \exp\left(-i E_a t\right), \qquad \rho(t)=\rho_0,
\qquad \cP_a(t)=1.
\eeq
More and more frequent measurements hinder the evolution and
eventually freeze it. This is the quantum Zeno paradox: ``A
watched pot never boils".

Notice that $\cV(t)$ in (\ref{eq:limita}) form a strongly
continuous one-parameter unitary group within $\cH_P$. Therefore,
starting from the dynamics (\ref{eq:VNAs}), which is irreversible
and probability-losing, one ends up with a fully reversible
evolution. Reversibility is recovered in the limit.

For one-dimensional projections, interesting behaviors can be
obtained (in the very limit $N\to\infty$) only by relaxing the
hypothesis $\ket{a}\in D(H)$. In such a case, the survival
probability, at variance with (\ref{eq:quadratic}), can be no more
quadratic at short times and one can obtain different results
depending on its short-time behavior.\cite{Muga} Assume, for
example, that
\beq
p_a(\tau)\sim 1-\left(\frac{\tau}{\tau_c}\right)^\alpha, \qquad
\mathrm{for} \quad \tau\to0,
\eeq
for some $\alpha\ge0$. The effective decay rate
(\ref{eq:effective}) becomes
\beq
\gamma(\tau)\sim \frac{\tau^{\alpha-1}}{\tau_c^\alpha},
\qquad\mathrm{for} \quad \tau\to0.
\eeq
When $\alpha> 1$,  $\gamma(0^+)=0$ and the limit (\ref{eq:limita})
is recovered again. On the other hand, for $0<\alpha<1$,
$\gamma(0^+)=+\infty$, and the limit operator $\cV(t)$ in Eq.
(\ref{eq:cVa}) vanishes. In such a case irreversibility
``survives" the limit and probability is \textit{immediately}
lost. Notice that in this case the limit operator is not
continuous at $t=0$, $\cV(0^+)=0\neq \cV(0)=P_a$, whence Misra and
Sudarshan's theorem does not apply. [Note: from (\ref{eq:VNAs}),
$\cV(0)=\lim_N V_N(0)=P_a$ .]

An interesting case arises at the threshold value $\alpha=1$, when
$\gamma(0^+)=1/\tau_c\neq 0$. Irreversibility still survives the
limit, but it manifests itself in a gentler way, through the limit
operator
\beq
\cV(t)=P_a \exp(- t/2\tau_c), \qquad \mathrm{for}\quad t\ge0 :
\label{eq:cVgamma}
\eeq
probability is not conserved and the Zeno paradox does not arise.
Notice that in Eq.\ (\ref{eq:cVgamma}) we had to assume
$\omega(0^+)<\infty$, in order to assure the existence of
$\cV(t)$. Finally, the limit case $\alpha=0$ trivially occurs when
the projection commutes with the Hamiltonian $[H,P_a]=0$, hence
$\ket{a}$ is an eigenstate of $H$ and does not evolve.

\subsection{Incomplete measurements}
 \label{sec-partial}
 \andy{partial}

In the case of incomplete measurements some outcomes are lumped
together, for example, because the experimental equipment has
insufficient resolution.\cite{Peres98} Therefore, the projection
operator $P$, which selects a particular lump, is
many-dimensional. Let us first consider a finite dimensional
$\cH_P=\mathrm{Ran}P$,
\beq
\mathrm{Tr} P=s, \qquad \mathrm{with}\quad s<\infty .
\eeq
The resulting time evolution operator is just a generalization of
(\ref{eq:VNAs}) to a finite dimensional matrix and has the
explicit form [see (\ref{eq:limita})]
\beq
\cV (t)= \lim_{N\to\infty} V_N(t) = \lim_{N\to\infty} [ P U(t/N) P
]^N =P \exp(-i P H P t).
\label{eq:cVfin1}
\eeq
As shown in Sec.\ \ref{sec-selmeas}, if $\cH_P\subset D(H)$, then
$\cV(t)$ in (\ref{eq:cVfin1}) is unitary within $\cH_P$ and is
generated by a resulting self-adjoint Hamiltonian
$PHP$.\cite{MNPRY} Again, reversibility is recovered in the limit.

For infinite dimensional projection, $s=\infty$, one can always
formally write the limiting evolution in the form
(\ref{eq:cVfin1}) but has to define the meaning of $P H P$. In
such a case the time evolution operator $\cV(t)$ may be not
unitary and interesting phenomena can arise, related to the
self-adjointness of the limiting Hamiltonian $PHP$. This an
interesting problem\cite{compactregularize,Friedman72} that will
not be discussed here.

In general, for incomplete selective measurements, the system Q
does not remain in its initial state, after the series of very
frequent measurements. Rather, the system is confined (and
evolves) in the subspace $\cH_P$, instead of evolving ``naturally"
in the total Hilbert space $\cH$.

\subsection{Nonselective measurements}
 \label{sec-nonselect}
 \andy{nonselect}

We will now consider the case of nonselective measurements and
extend Misra and Sudarshan's theorem\cite{Misra} in order to
accomodate multiple projectors and to build a bridge for our
subsequent discussion. For nonselective measurements the measuring
apparatus functions, but no selection of outcomes is
performed,\cite{Schwinger59} and all ``beams" go through the whole
Zeno dynamics. Let $\{P_n\}_n$,
\beq
P_nP_m=\delta_{mn}P_n,\qquad  \sum_n P_n=1 ,
\eeq
be a (countable) collection of projection operators and
$\mathrm{Ran}P_n=\cH_{P_n}$ the relative subspaces. This induces a
partition on the total Hilbert space
\beq
\cH=\bigoplus_n \cH_{P_n}.
\eeq
Consider the associated nonselective measurement described by the
superoperator
\beq
\label{eq:superP}
\hat P \rho=\sum_n P_n \rho P_n.
\eeq
The free evolution reads
\beq
\hat U_t \rho_0=U(t) \rho_0 U^\dagger(t),\qquad U(t)=\exp(-i H t)
\eeq
and the Zeno evolution after $N$ measurements in a time $t$ is
governed by the superoperator
\beq
\hat V^{(N)}_t=\hat P\left(\hat U \left(t/N \right)\hat
P\right)^{N-1} ,
\eeq
which yields
\beq
\rho(t)=\hat V^{(N)}_t \rho_0 =\sum_{n_1,\dots,n_N}V_{n_1\dots
n_N}^{(N)}(t)\; \rho_0\; V_{n_1\dots n_N}^{(N)\dagger}(t) ,
\eeq
where
\barr
V_{n_1\dots n_N}^{(N)}(t)  = P_{n_N} U\left(t/N\right) P_{n_{N-1}}
\cdots P_{n_2} U\left(t/N\right) P_{n_1}.
\label{eq:boo}
\earr
We follow Misra and Sudarshan\cite{Misra} and assume, as in Sec.\
\ref{sec-MisSud}, the existence of the strong limits ($t>0$)
\andy{slims}
\beq
\cV_n (t)=\lim_{N\to\infty} V_{n\dots n}^{(N)}(t) , \qquad \lim_{t
\rightarrow 0^+} \cV_n(t) = P_n , \quad \forall n \ .
\label{eq:slims}
\eeq
Then $\cV_n(t)$ exist for all real $t$ and form a
semigroup,\cite{Misra} and
\beq
\cV_n^\dagger(t)\cV_n(t)=P_n.
\eeq
Moreover, it is easy to show that
\beq
\lim_{N\to\infty} V_{n\dots n'\dots}^{(N)}(t) = 0, \qquad
\mathrm{for}\quad n'\neq n .
\eeq
Therefore the final state is
\andy{rhoZ}
\barr
\rho(t)=\hat \cV_t\rho_0 =\sum_n \cV_n(t) \rho_0 \cV_n^\dagger(t),
\quad \mathrm{with} \quad \sum_n \cV_n^\dagger(t)\cV_n(t)=\sum_n
P_n=1 .\;\;
\label{eq:rhoZ}
\earr
The components $\cV_n(t) \rho_0 \cV_n^\dagger(t)$ make up a block
diagonal matrix: the initial density matrix is reduced to a
mixture and any interference between different subspaces
$\cH_{P_n}$ is destroyed (complete decoherence). In conclusion,
\andy{probinfu}
\beq
p_n(t) =  \mathrm{Tr} \left[\rho(t) P_n\right]=
\mathrm{Tr}\left[\rho_0 P_n\right]=p_n(0) , \quad \forall n .
\label{eq:probinfu}
\eeq
In words, probability is conserved in each subspace and no
probability ``leakage" between any two subspaces is possible. The
total Hilbert space splits into invariant subspaces and the
different components of the wave function (or density matrix)
evolve independently within each sector. One can think of the
total Hilbert space as the shell of a tortoise, each invariant
subspace being one of the scales. Motion among different scales is
impossible. (See Fig.\ \ref{tortoise} in the following.)

If $\mathrm{Tr} P_n=s_n<\infty$, then the limiting evolution
operator $\cV_n(t)$ (\ref{eq:slims}) within the subspace
$\cH_{P_n}$ has the form (\ref{eq:cVfin1}),
\beq
\cV_n (t)=P_n \exp(-i P_n H P_n t) ,
\label{eq:cVfin}
\eeq
is unitary in $\cH_{P_n}$ and the resulting Hamiltonian $P_n H
P_n$ is self-adjoint, provided that $\cH_{P_n}\subset D (H)$.

The original limit result (\ref{eq:probinfu1}) is reobtained when
$p_n=1$ for some $n$, in (\ref{eq:probinfu}): the initial state is
then in one of the invariant subspaces and the survival
probability in that subspace remains unity. However, even if the
limits are the same, notice that the setup described here is
conceptually different from that of Sec.\ \ref{sec-MisSud}.
Indeed, the dynamics (\ref{eq:boo}) allows transitions among
different subspaces $\cH_{P_n}\to\cH_{P_m}$, while the dynamics
(\ref{eq:Nproie}) completely forbids them. Therefore, for finite
$N$, (\ref{eq:boo}) takes into account the possibility that one
subspace $\cH_{P_n}$ gets
\textit{repopulated}\cite{Nakazato96b,zenoreview} after the system
has made transitions to other subspaces, while in
(\ref{eq:Nproie}) the system must be found in $\cH_{P_n}$ at every
measurement.

\section{Dynamical quantum Zeno effect}
 \label{sec-dynamicalQZE}
 \andy{dynamicalQZE}

All our discussion has dealt so far with ``pulsed" measurements,
according to von Neumann's projection postulate.\cite{von}
However, from a physical point of view, a ``measurement" is
nothing but an interaction with an external system (another
quantum object, or a field, or simply another degree of freedom of
the very system investigated), playing the role of apparatus. We
emphasize that in such a case the QZE is a consequence of the
dynamical features (i.e.\ the form factors) of the coupling
between the system investigated and the external system, and no
use is made of projection operators (and non-unitary dynamics).
The idea of ``continuous" measurement in a QZE context has been
proposed several times during the last two
decades,\cite{Itano,Peres80,PeresKraus} although the first
quantitative comparison with the ``pulsed" situation is rather
recent.\cite{Schulman98}

We consider therefore a purely dynamical evolution, by including
the detector in the quantum description. In
general,\cite{zenoreview,Napoli} one can consider the Hamiltonian
\beq
\label{eq:sys+meas}
H_K=H+ K H_{\mathrm{meas}},
\eeq
where $H$ is the Hamiltonian of the system under observation (and
can include the free Hamiltonian of the apparatus,
$H=H_{\mathrm{sys}}+H_{\mathrm{det}}$) and $H_{\mathrm{meas}}$ is
the interaction Hamiltonian between the system and the apparatus,
$K$ representing the strength of the measurement or, equivalently,
the inverse response time of the apparatus [see examples in Sec.\
(\ref{sec-applications})].

\subsection{A theorem}
 \label{sec-dyntheo}
 \andy{dyntheo}

We now state a theorem,\cite{thesis,theorem} which is the exact
analog of Misra and Sudarshan's theorem for a dynamical evolution
of the type (\ref{eq:sys+meas}). Consider the time evolution
operator
\barr
U_{K}(t) = \exp(-iH_K t) .
\label{eq:measinter}
\earr
We will prove that in the ``infinitely strong measurement"
(``infinitely quick detector") limit $K\to\infty$ the evolution
operator
\beq
\label{eq:limevol}
\cU(t)=\lim_{K\to\infty}U_{K}(t),
\eeq
becomes diagonal with respect to $H_{\mathrm{meas}}$:
\beq
\label{eq:diagevol}
[\cU(t), P_n]=0, \qquad
\mathrm{where}
\quad
H_{\mathrm{meas}} P_n=\eta_n P_n,
\eeq
$P_n$ being the orthogonal projection onto $\cH_{P_n}$, the
eigenspace of $H_{\mathrm{meas}}$ belonging to the eigenvalue
$\eta_n$. Note that in Eq.\ (\ref{eq:diagevol}) one has to
consider distinct eigenvalues, i.e., $\eta_n\neq\eta_m$ for $n\neq
m$, whence the $\cH_{P_n}$'s are in general multidimensional.

Moreover, the limiting evolution operator has the explicit form
\beq
\label{eq:theorem}
\cU(t)=\exp[-i(H^{\mathrm{diag}}+K H_{\mathrm{meas}}) t],
\eeq
where
\beq
H^{\mathrm{diag}}=\sum_n P_n H P_n
\label{eq:diagsys}
\eeq
is the diagonal part of the system Hamiltonian $H$ with respect to
the interaction Hamiltonian $H_{\mathrm{meas}}$.

\subsection{Dynamical superselection rules}
 \label{sec-supersel}
 \andy{supersel}

\begin{figure}[t]
%\figurebox{20pc}{15pc}{} % to have a box alone
\epsfxsize=11pc % will enlarge or reduce the postscript figures based on the xsize
\centerline{\epsfbox{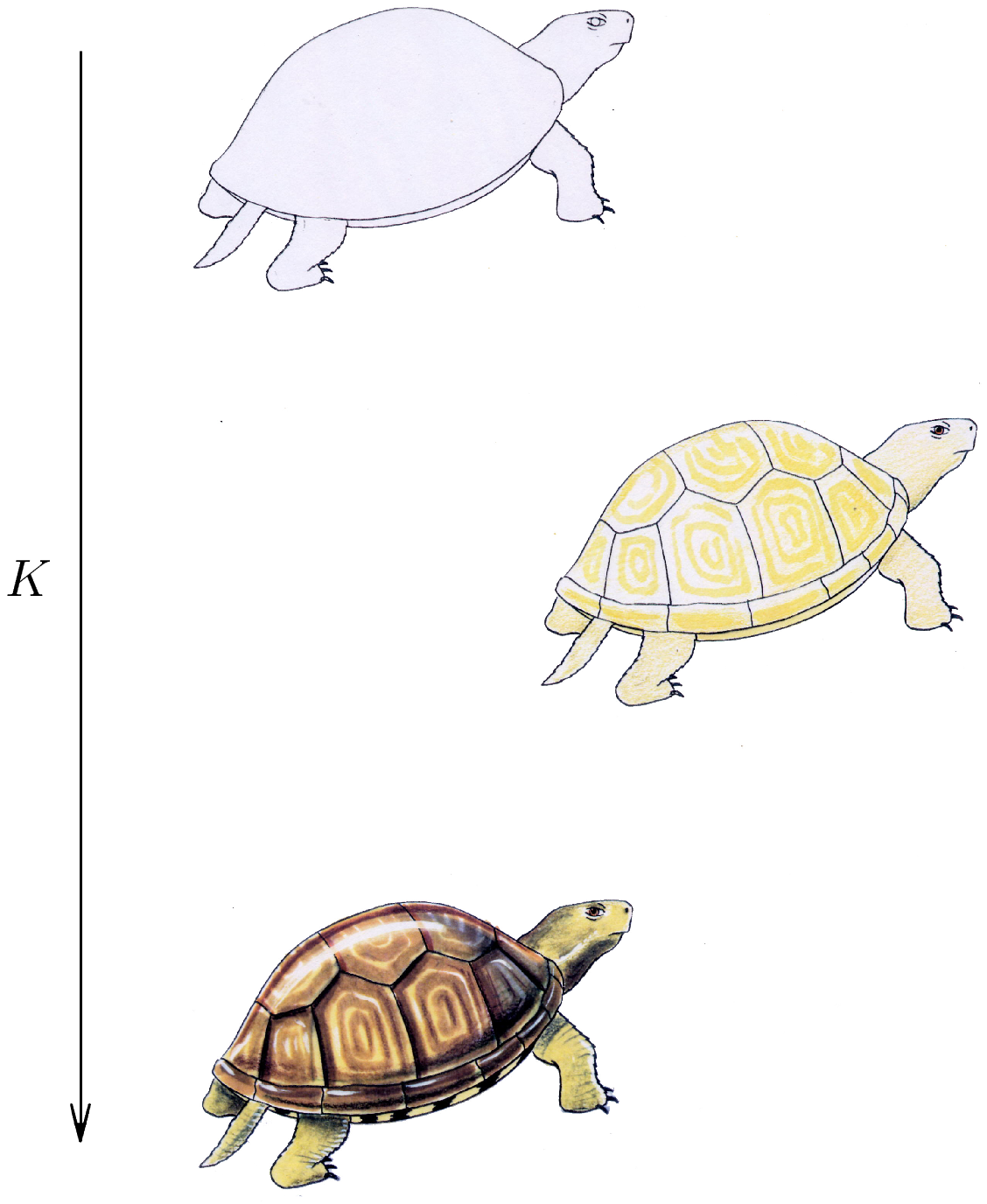}} % postscript image file name
\caption{\label{tortoise} The Hilbert space of the system:
a superselection rule appears as the coupling $K$ to the apparatus
is increased.}
\end{figure}

Before proving the theorem of Sec.\ \ref{sec-dyntheo} let us
briefly consider its physical implications. In the $K\to\infty$
limit, due to (\ref{eq:diagevol}), the time evolution operator
becomes diagonal with respect to $H_{\mathrm{meas}}$, namely
\beq
[\cU(t), H_{\mathrm{meas}}]=0,
\eeq
a superselection rule arises and the total Hilbert space is split
into subspaces $\cH_{P_n}$ which are invariant under the
evolution. These subspaces are simply defined by the $P_n$'s,
i.e., they are eigenspaces belonging to distinct eigenvalues
$\eta_n$: in other words, \textit{subspaces that the apparatus is
able to distinguish}. On the other hand, due to
(\ref{eq:diagsys}), the dynamics within each Zeno subspace
$\cH_{P_n}$ is governed by the diagonal part $P_n H P_n$ of the
system Hamiltonian $H$. The evolution reads
\barr
\rho(t)= \cU(t) \rho_0 \cU^\dagger(t) =e^{-i (H^{\mathrm{diag}}+K
H_{\mathrm{meas}}) t}\rho_0 e^{i (H^{\mathrm{diag}}+K
H_{\mathrm{meas}}) t}
\earr
and the probability to find the system in $\cH_{P_n}$
\barr
p_n(t)&=&\mathrm{Tr} \left[ \rho(t) P_n \right]= \mathrm{Tr}
\left[\cU(t)\rho_0\cU^\dagger(t) P_n\right] =\mathrm{Tr}
\left[\cU(t)\rho_0 P_n\cU^\dagger(t)\right]
\nonumber\\
&=& \mathrm{Tr} \left[ \rho_0 P_n \right]=p_n(0)
\label{eq:pntpn0}
\earr
is constant. As a consequence, if the initial state of the system
belongs to a specific sector, it will be forced to remain there
forever (QZE):
\beq
\psi_0\in \cH_{P_n}\rightarrow \psi(t)\in \cH_{P_n}.
\eeq
More generally, if the initial state is an incoherent
superposition of the form $\rho_0=\hat P \rho_0$, with $\hat P$
defined in (\ref{eq:superP}), then each component will evolve
separately, according to
\barr
\rho(t)&=&\cU(t)\rho_0\cU^\dagger(t)=\sum_n e^{-i
(H^{\mathrm{diag}}+K H_{\mathrm{meas}})t} P_n\rho_0 P_n e^{i
(H^{\mathrm{diag}}+K H_{\mathrm{meas}}) t}
\nonumber\\
&=&\sum_n e^{-i P_n H P_n t} P_n\rho_0 P_n e^{i P_n H P_n t}=
\sum_n \cV_n(t) \rho_0 \cV_n^\dagger(t),
\earr
with $\cV_n(t)= P_n \exp(-i P_n H P_n t)$, which is exactly the
same result (\ref{eq:rhoZ})-(\ref{eq:cVfin}) found for the case of
nonselective pulsed measurements.  This bridges the gap with the
description of Sec.\ \ref{sec-nonselect} and clarifies the role of
the detection apparatus. In Fig.\
\ref{tortoise} we endeavored to give a pictorial representation of
the decomposition of the Hilbert space as $K$ is increased.

\subsection{Proof of the theorem}
 \label{sec-proof}
 \andy{proof}

We will now use perturbation theory and prove\cite{thesis} that
the limiting evolution operator has the form (\ref{eq:theorem}).
From that, property (\ref{eq:diagevol}) follows. In the next
section we will give a more direct proof of (\ref{eq:diagevol}),
which relies on the adiabatic theorem.

Rewrite the time evolution operator in the form
\beq
\label{eq:timevoltau}
U_K(t)=\exp(-i H_K t)=\exp(-i H_\lambda \tau)=U_\lambda(\tau)
\eeq
where
\beq
\lambda=1/K,\qquad \tau= K t=t/\lambda, \qquad H_\lambda=\lambda
H_K= H_{\mathrm{meas}}+\lambda H,
\label{eq:lambdaK}
\eeq
and apply  perturbation theory to the Hamiltonian $H_\lambda$ for
small $\lambda$. To this end choose the unperturbed degenerate
projections $P_{n\alpha}$
\beq
H_{\mathrm{meas}}P_{n\alpha}=\eta_n P_{n\alpha}, \qquad P_n=
\sum_{\alpha}P_{n\alpha},
\eeq
whose degeneration $\alpha$ is resolved at some order in the
coupling constant $\lambda$. This means that by denoting with
 $\widetilde \eta_{n\alpha}$ and $\widetilde P_{n\alpha}$ the eigenvalues and
 the orthogonal projections of the total Hamiltonian $H_{\lambda}$
\beq
H_\lambda\widetilde{P}_{n\alpha}=\widetilde{\eta}_{n\alpha}\widetilde{P}_{n\alpha},
\eeq
they reduce to the unperturbed ones when the perturbation vanishes
\beq
\widetilde{P}_{n\alpha}\stackrel{\lambda\to
0}{\longrightarrow}P_{n\alpha},\qquad
\widetilde{\eta}_{n\alpha}\stackrel{\lambda\to
0}{\longrightarrow}\eta_n.
\eeq
Therefore, by applying standard perturbation
theory,\cite{Messiah61} we get the eigenvectors
\barr
\widetilde{P}_{n\alpha}&=&P_{n\alpha}+\lambda
P_{n\alpha}^{(1)}+\Ord(\lambda^2)
\nonumber\\
&=&P_{n\alpha}+\lambda \left( \frac{Q_n}{a_n}H
P_{n\alpha}+P_{n\alpha}H\frac{Q_n}{a_n}\right)+\Ord(\lambda^2) ,
\label{eq:perteigenvec}
\earr
where
\barr
Q_n=1-P_n=\sum_{m\neq n}P_{m}, \qquad
\frac{Q_n}{a_n}=\frac{Q_n}{\eta_n-H_{\mathrm{meas}}}=\sum_{m\neq
n}\frac{P_m}{\eta_n-\eta_m} .
\earr
The perturbative expansion of the eigenvalues reads
\barr
\widetilde{\eta}_{n\alpha}=\eta_n+\lambda
\eta_{n\alpha}^{(1)}+\lambda^2 \eta_{n\alpha}^{(2)}+\Ord(\lambda^3)
\label{eq:perteigenval}
\earr
where
\barr
\eta_{n\alpha}^{(1)} P_{n\alpha} &=& P_{n\alpha} H P_{n\alpha},
\qquad
\eta_{n\alpha}^{(2)} P_{n\alpha} = P_{n\alpha} H \frac{Q_n}{a_n} H
P_{n\alpha},
\nonumber\\
P_{n\alpha} H P_{n\beta} &=& P_{n\alpha} H \frac{Q_n}{a_n} H
P_{n\beta}=0,\qquad \alpha\neq\beta .
\label{eq:perteigenval1}
\earr
 Write now the spectral decomposition of the evolution operator\
(\ref{eq:timevoltau}) in terms of the projections $\widetilde
P_{n\alpha}$
\beq
U_\lambda(\tau)=\exp(-iH_\lambda
\tau)\sum_{n,\alpha}\widetilde{P}_{n\alpha}=
\sum_{n,\alpha}\exp(-i\widetilde{\eta}_{n\alpha}
\tau)\widetilde{P}_{n\alpha}
\eeq
and plug in the perturbation expansions\ (\ref{eq:perteigenvec}),
to obtain
\barr
U_\lambda(\tau)
&=&\sum_{n,\alpha}e^{-i\widetilde{\eta}_{n\alpha}\tau}P_{n\alpha}
\nonumber\\
& & +\lambda\sum_{n,\alpha}\left(\frac{Q_n}{a_n}H P_{n\alpha}
e^{-i\widetilde{\eta}_{n\alpha}\tau}
+e^{-i\widetilde{\eta}_{n\alpha}\tau}P_{n\alpha}H\frac{Q_n}{a_n}
\right)+\Ord(\lambda^2).
\label{eq:Upertexp}
\earr
Let us define a new operator $\widetilde{H}_{\lambda}$ as
\barr
\widetilde{H}_{\lambda}&=&\sum_{n,\alpha}\widetilde{\eta}_{n\alpha}P_{n\alpha}
\nonumber\\
&=&H_{\mathrm{meas}}+\lambda\sum_n P_n H P_n+ \lambda^2\sum_n P_n
H \frac{Q_n}{a_n}H P_n+\Ord(\lambda^3),
\label{eq:tildeHl}
\earr
where Eqs.\ (\ref{eq:perteigenval})-(\ref{eq:perteigenval1}) were
used. By plugging Eq.\ (\ref{eq:tildeHl}) into Eq.\
(\ref{eq:Upertexp}) and making use of the property
\beq
\sum_n P_n H \frac{Q_n}{a_n}=-\sum_n\frac{Q_n}{a_n}H P_{n} ,
\eeq
we finally obtain
\barr
U_\lambda(\tau) =\exp(-i\widetilde{H}_\lambda \tau)+
\lambda\left[ \sum_n\frac{Q_n}{a_n}H P_{n}, \;
\exp(-i\widetilde{H}_\lambda \tau)\right] +\Ord(\lambda^2),
\label{eq:Upertexp1}
\earr
Now, by recalling the definition\ (\ref{eq:lambdaK}), we can write
the time evolution operator $U_K(t)$ as the sum of two terms
\beq
U_K(t)=U_{\mathrm{ad}, K}(t)+\frac{1}{K} U_{\mathrm{na},K}(t),
\eeq
where
\barr
U_{\mathrm{ad},K}(t)=e^{-i\left( K H_{\mathrm{meas}}+\sum_n P_n H
P_n+
\frac{1}{K}\sum_n P_n H \frac{Q_n}{a_n}H
P_n+\Ord\left(K^{-2}\right)\right)t} \;\;\;\;
\earr
is a diagonal, \textit{adiabatic} evolution and
\beq
U_{\mathrm{na},K}(t)=\left[\sum_n
\frac{Q_n}{a_n}H P_n,\; U_{\mathrm{ad},K}(t)\right]+\Ord\left(K^{-1}\right)
\eeq
is the off-diagonal, \textit{nonadiabatic} correction. In the
$K\to\infty$ limit only the adiabatic term survives and one
obtains
\beq
\cU(t)=\lim_{K\to\infty}U_{K}(t)=\lim_{K\to\infty}
U_{\mathrm{ad},K}(t)=e^{-i \left(K H_{\mathrm{meas}}+\sum_n P_n H
P_n\right) t},
\eeq
which is formula\ (\ref{eq:theorem}) [and implies also
(\ref{eq:diagevol})]. The proof is complete. As a byproduct we get
the corrections to the exact limit, valid for large, but finite,
values of $K$.

\subsection{Zeno evolution from an adiabatic theorem}
 \label{sec-Zenoadiab}
 \andy{Zenoadiab}

We now give an alternative proof [and a generalization to
time-dependent Hamiltonians $H(t)$] of Eq.\ (\ref{eq:diagevol}).
The adiabatic theorem deals with the time evolution operator
$U(t)$ when the Hamiltonian $H(t)$ depends slowly on time. The
traditional formulation\cite{Messiah61} replaces the physical time
$t$ by the scaled time $s=t/T$ and considers the solution of the
scaled Schr\"odinger equation
\beq
i \frac{d}{ds} U_T (s) = T H(s) U_T (s)
\label{eq:adiabatic}
\eeq
in the $T\to\infty$ limit.

Given a family  $P(s)$ of smooth spectral projections of $H(s)$
\beq
H(s)P(s)=E(s)P(s),
\eeq
the adiabatic time evolution $U_{\mathrm{A}}(s)=\lim_{T\to\infty}
U_T(s)$ has the intertwining property\cite{adiabatic,Messiah61}
\beq
U_{\mathrm{A}}(s)P(0)=P(s) U_{\mathrm{A}} (s) ,
\label{eq:adintert}
\eeq
that is, $U_{\mathrm{A}}(s)$ maps $\cH_{P(0)}$ onto $\cH_{P(s)}$.

Theorem (\ref{eq:diagevol}) and its generalization,
\beq
\cU(t) P_n(0)=P_n(t) \cU(t),
\label{eq:theoremt}
\eeq
valid for generic time dependent Hamiltonians,
\beq
\label{eq:sys+meast}
H_K(t)=H(t)+ K H_{\mathrm{meas}}(t),
\eeq
are easily proven by recasting them in the form of an adiabatic
theorem.\cite{theorem} In the $H$ interaction picture, given by
\beq
i \frac{d}{dt} U_{\mathrm{S}}(t) = H U_{\mathrm{S}}(t), \qquad
H^{\mathrm{I}}_{\mathrm{meas}}(t)=U^\dagger_{\mathrm{S}}(t)
H_{\mathrm{meas}} U_{\mathrm{S}}(t),
\eeq
the Schr\"odinger equation reads
\beq
i \frac{d}{dt} U_K^{\mathrm{I}} (t) = K
H^{\mathrm{I}}_{\mathrm{meas}}(t) \; U_K^{\mathrm{I}} (t).
\label{eq:adiabaticlike}
\eeq
The Zeno evolution pertains to the $K\to\infty$ limit. And in such
a limit Eq.\ (\ref{eq:adiabaticlike}) has exactly the same form of
the adiabatic evolution (\ref{eq:adiabatic}): the large coupling
$K$ limit corresponds to the large time $T$ limit and the physical
time $t$ to the scaled time $s=t/T$. Therefore, let us consider a
spectral projection of $H^{\mathrm{I}}_{\mathrm{meas}}(t)$,
\beq
P^{\mathrm{I}}_n(t)=U^\dagger_{\mathrm{S}}(t) P_n(t)
U_{\mathrm{S}}(t) ,
\label{eq:PIn}
\eeq
such that
\beq
H^{\mathrm{I}}_{\mathrm{meas}}(t) P^{\mathrm{I}}_n(t)= \eta_n(t)
P^{\mathrm{I}}_n(t), \qquad H_{\mathrm{meas}}(t) P_n(t)= \eta_n(t)
P_n(t) .
\eeq
The limiting operator
\beq
\cU^{\mathrm{I}} (t)=\lim_{K\to\infty}U_K^{\mathrm{I}}(t)
\eeq
has the intertwining property (\ref{eq:adintert})
\beq
\cU^{\mathrm{I}} (t)
P^{\mathrm{I}}_n(0)=P^{\mathrm{I}}_n(t)\cU^{\mathrm{I}} (t),
\eeq
i.e.\ maps $\cH_{P^{\mathrm{I}}_n(0)}$ onto
$\cH_{P^{\mathrm{I}}_n(t)}$:
\beq
\psi^{\mathrm{I}}_0\in \cH_{P^{\mathrm{I}}_n(0)}\rightarrow
\psi^{\mathrm{I}}(t)\in \cH_{P^{\mathrm{I}}_n(t)}.
\eeq
In the Schr\"odinger picture the limiting operator
\beq
\cU(t)=\lim_{K\to\infty} U_{\mathrm{S}}(t)
U_K^{\mathrm{I}}(t)=U_{\mathrm{S}}(t)\cU^{\mathrm{I}}(t)
\eeq
satisfies the intertwining property (\ref{eq:theoremt}) [see
(\ref{eq:PIn})]
\barr
\cU(t) P_n(0) &=& U_{\mathrm{S}}(t)\cU^{\mathrm{I}}(t) P_n(0)
=U_{\mathrm{S}}(t)\cU^{\mathrm{I}} (t)
P^{\mathrm{I}}_n(0)\nonumber\\
&=&U_{\mathrm{S}}(t) P^{\mathrm{I}}_n(t)\cU^{\mathrm{I}} (t) =
P_n(t) U_{\mathrm{S}}(t) \cU^{\mathrm{I}}(t)=P_n(t) \cU(t),
\label{eq:intertwin}
\earr
and maps $\cH_{P_n(0)}$ onto $\cH_{P_n(t)}$:
\beq
\psi_0\in \cH_{P_n(0)}\rightarrow \psi(t)\in \cH_{P_n(t)}.
\label{eq:mapS}
\eeq
The probability to find the system in $\cH_{P_n(t)}$,
\barr
p_n(t)&=&\mathrm{Tr}\left[P_n(t) \cU(t) \rho_0
\cU^\dagger(t)\right] =\mathrm{Tr}\left[\cU(t) P_n(0)\rho_0
\cU^\dagger(t)\right]\nonumber\\
&=&\mathrm{Tr}\left[P_n(0)\rho_0 \right]=p_n(0),
\label{eq:pntpn01}
\earr
is constant: if the initial state of the system belongs to a given
sector, it will be forced to remain there forever (QZE).

For a time independent Hamiltonian
$H_{\mathrm{meas}}(t)=H_{\mathrm{meas}}$, the projections are
constant, $P_n(t)=P_n$, hence Eq.\ (\ref{eq:theoremt}) reduces to
(\ref{eq:diagevol}) and the above property holds
\textit{a fortiori} and reduces to (\ref{eq:pntpn0}).

\subsection{Generalizations}
 \label{sec-general}
 \andy{general}

The formulation of a Zeno dynamics in terms of an adiabatic
theorem is powerful. Indeed one can use all the machinery of
adiabatic theorems in order to get results in this context. An
interesting extension would be to consider time-dependent
measurements
\beq
H_{\mathrm{meas}}=H_{\mathrm{meas}}(t),
\eeq
whose spectral projections $P_n=P_n(t)$ have a nontrivial time
evolution. In this case, instead of confining the quantum state to
a fixed sector, one can transport it along a given path (subspace)
$\cH_{P_n(t)}$, according to Eqs.\
(\ref{eq:mapS})-(\ref{eq:pntpn01}). One then obtains a dynamical
generalization of the process pioneered by Von Neumann in terms of
projection operators.\cite{von,AV}

\section{Applications}
 \label{sec-applications}
 \andy{applications}
As a first example, consider the Hamiltonian
\andy{ham3l}
\beq
H_{\mathrm{3lev}} = H+ K H_{\mathrm{meas}}= \pmatrix{0 & \Omega &
0\cr
\Omega & 0 & K \cr 0 & K & 0},
\label{ham3l}
\eeq
describing a two-level system, with Hamiltonian
\beq
H=\Omega (\ket{1}\bra{2}+ \ket{2}\bra{1})=\Omega
\pmatrix{
  0 & 1 & 0 \cr
  1 & 0 & 0 \cr
  0 & 0 & 0
},
\eeq
coupled to a third one, that plays the role of measuring
apparatus:
\beq
H_{\mathrm{ meas}}= \ket{2}\bra{3}+\ket{3}\bra{2} =
\pmatrix{
  0 & 0 & 0 \cr
  0 & 0 & 1 \cr
  0 & 1 & 0
}.
\label{eq:Hmeas2}
\eeq
This example was considered by Peres.\cite{Peres80} One expects
the third level  to perform better as a measuring apparatus when
the coupling $K$ becomes larger. Indeed, if initially the system
is in state $\ket{1}$, the survival probability
reads\cite{zenoreview}
\andy{sp3}
\beq
p_0(t)= \frac{1}{K_1^4} \left[K^2+ \Omega^2 \cos(K_1 t)
\right]^2\; \stackrel{K\to \infty}{\longrightarrow} 1, \qquad
K_1=\sqrt{K^2+\Omega^2} \ .
\label{sp3}
\eeq
In spite of its simplicity, this model clarifies the physical
meaning of a ``continuous" measurement performed by an ``external
apparatus" (which can even be another degree of freedom of the
system investigated). Also, it captures many interesting features
of a Zeno dynamics. Indeed, as $K$ is increased, the Hilbert space
is split into three invariant subspaces $\cH=\bigoplus\cH_{P_n}$,
the three eigenspaces of $H_{\mathrm{meas}}$:
\beq
\cH_{P_0}=\{ \ket{1}\}, \quad \cH_{P_1}=\{
(\ket{2}+\ket{3})/\sqrt{2}\}, \quad \cH_{P_{-1}}=\{
(\ket{2}-\ket{3})/\sqrt{2}\},
\label{eq:3sub}
\eeq
corresponding to projections
\beq
P_0=\pmatrix{ 1 & 0 & 0 \cr 0 & 0 & 0 \cr 0 & 0 & 0 }, \quad
P_1=\frac{1}{2}\pmatrix{ 0 & 0 & 0 \cr 0 & 1 & 1 \cr 0 & 1 & 1 },
\quad P_{-1}=\frac{1}{2}\left(\begin{array}{rrr}
  0 & 0 & 0 \\
  0 & 1 & -1 \\
  0 & -1 & 1
\end{array}\right),
\eeq
with eigenvalues $\eta_0=0$ and $\eta_{\pm1}=\pm1$. Therefore the
diagonal part of the system Hamiltonian $H$ vanishes,
$H^{\mathrm{diag}}=\sum P_n H P_n=0$, the Zeno evolution is
governed by
\beq
H^{\mathrm{diag}}+K H_{\mathrm{meas}}=\pmatrix{0 & 0 & 0\cr 0 & 0
& K \cr 0 & K & 0}
\eeq
and any transition between $\ket{1}$ and $\ket{2}$ is inhibited: a
watched pot never boils.

Second example: consider
\andy{ham4l}
\beq
H_{\mathrm{4lev}} =\Omega \sigma_1+ K\tau_1+K' \tau'_1 =
\pmatrix{0 & \Omega & 0 & 0 \cr \Omega & 0 & K & 0 \cr 0 & K & 0 &
K' \cr 0 & 0 & K' & 0 },
\label{ham4l}
\eeq
where states $\ket{1}$ and $\ket{2}$ make Rabi oscillations,
\beq
\Omega \sigma_1=\Omega (\ket{2}\bra{1}+\ket{1}\bra{2})= \Omega
\pmatrix{0 & 1 & 0 & 0 \cr 1 & 0 & 0 & 0 \cr 0 & 0 & 0 & 0 \cr 0 &
0 & 0 & 0 } ,
\eeq
while state $\ket{3}$ ``observes" them,
\beq
K \tau_1=K (\ket{3}\bra{2}+\ket{2}\bra{3})= K \pmatrix{0 & 0& 0 &
0 \cr 0 & 0 & 1 & 0 \cr 0 & 1 & 0 & 0 \cr 0 & 0 & 0 & 0 } ,
\eeq
and state $\ket{4}$ ``observes" whether level $\ket{3}$ is
populated,
\beq
K' \tau'_1=K' (\ket{4}\bra{3}+\ket{3}\bra{4})= K' \pmatrix{0 & 0&
0 & 0 \cr 0 & 0 & 0 & 0 \cr 0 & 0 & 0 & 1 \cr 0 & 0 & 1 & 0 } .
\eeq
If $K \gg \Omega$ \textit{and} $K'$, then the total Hilbert space
is split into the three eigenspaces of $\tau_1$ [compare with
(\ref{eq:3sub})]:
\beq
\cH_{P_0}=\{ \ket{1}, \ket{4}\}, \quad \cH_{P_1}=\{
(\ket{2}+\ket{3})/\sqrt{2}\}, \quad \cH_{P_{-1}}=\{
(\ket{2}-\ket{3})/\sqrt{2}\},
\label{eq:3sub1}
\eeq
the Zeno evolution is governed by
\beq
H_{\mathrm{4lev}}^{\mathrm{diag}}= K \tau_1=\pmatrix{0 & 0 & 0 & 0
\cr 0 & 0 & K & 0 \cr 0 & K & 0 & 0 \cr 0 & 0 & 0 & 0 }
\eeq
and the Rabi oscillations between states $\ket{1}$ and $\ket{2}$
are hindered. On the other hand, if $K' \gg K \gg \Omega$, the
total Hilbert space is instead divided into the three eigenspaces
of $\tau'_1$ [notice the differences with (\ref{eq:3sub1})]:
\beq
\cH_{P'_0}=\{ \ket{1}, \ket{2}\}, \quad \cH_{P'_1}=\{
(\ket{3}+\ket{4})/\sqrt{2}\}, \quad \cH_{P'_{-1}}=\{
(\ket{3}-\ket{4})/\sqrt{2}\},
\label{eq:3sub11}
\eeq
the Zeno Hamiltonian reads
\beq
H_{\mathrm{4lev}}^{\mathrm{diag}\prime}=
\Omega\sigma_1+K'\tau'_1=\pmatrix{0 &
\Omega & 0 & 0 \cr \Omega & 0 & 0 & 0 \cr 0 & 0 & 0 & K' \cr 0 & 0
& K' & 0 }
\eeq
and the $\Omega$ oscillations are fully \textit{restored} (in
spite of $K \gg \Omega$).\cite{Militello01} A watched cook can
freely watch a boiling pot.

Third example (decoherence-free subspaces\cite{Viola99} in quantum
computation). The Hamiltonian\cite{Beige00}
\beq
\label{eq:cavity}
H_{\mathrm{meas}}=i g \sum_{i=1}^2 \left( b\; \ket{2}_{ii}\bra{1}
- b^\dagger\; \ket{1}_{ii}\bra{2}\right) - i \kappa b^\dagger b
\eeq
describes a system of two ($i=1,2$) three-level atoms in a cavity.
The atoms are in a $\Lambda$ configuration with split ground
states $\ket{0}_i$ and $\ket{1}_i$ and excited state $\ket{2}_i$,
while the cavity has a single resonator mode $b$ in resonance with
the atomic transition 1-2. Spontaneous emission inside the cavity
is neglected, but a photon leaks out through the nonideal mirrors
with a rate $\kappa$.

The excitation number $\cN$,
\beq
\cN=\sum_{i=1,2}\ket{2}_{ii}\bra{2}+b^\dagger b ,
\eeq
commutes with the Hamiltonian,
\beq
[H_{\mathrm{meas}},\cN]=0.
\eeq
Therefore we can solve the eigenvalue equation inside each
eigenspace of $\cN$.

A comment is now in order. Strictly speaking, the Hamiltonian
(\ref{eq:cavity}) is non-Hermitian and we can not apply directly
the theorem of Sec.\ \ref{sec-dyntheo}. (Notice that the proof of
the theorem heavily hinges upon Hermiticity of Hamiltonians and
unitarity of evolutions.) However, we can enlarge our Hilbert
space $\cH$, by including the photon modes outside the cavity
$a_\omega$ and their coupling with the cavity mode $b$. The
enlarged dynamics is generated by the
\textit{Hermitian} Hamiltonian
\barr
\tilde H_{\mathrm{meas}} &=& i g \sum_{i=1}^2 \left( b\;
\ket{2}_{ii}\bra{1} - b^\dagger\; \ket{1}_{ii}\bra{2}\right)
\nonumber\\
& &+
\int d\omega\; \omega a^\dagger_\omega a_\omega +
\sqrt{\frac{\kappa}{\pi}}\int d\omega\left[a^\dagger_\omega b +
a_\omega b^\dagger\right].
\earr
It is easy to show that the evolution engendered by $\tilde
H_{\mathrm{meas}}$, when projected back to $\cH$, is given by the
effective non-Hermitian Hamiltonian (\ref{eq:cavity}), provided
the field outside the cavity is initially in the vacuum state.
Notice that any complex eigenvalue of $H_{\mathrm{meas}}$
engenders a dissipation of $\cH$ into the enlarged Hilbert space
embedding it. On the other hand, any real eigenvalue of
$H_{\mathrm{meas}}$ generates a unitary dynamics which preserves
probability within $\cH$. Hence it is also an eigenvalue of
$\tilde H_{\mathrm{meas}}$ and its eigenvectors are the
eigenvectors of the restriction $\tilde H_{\mathrm{meas}}|_{\cH}$.
Therefore, as a general rule, the theorem of Sec.\
\ref{sec-dyntheo} can be applied also to non-Hermitian measurement
Hamiltonians $\cH_{\mathrm{meas}}$, provided one restricts one's
attention only to their real eigenvalues.

The eigenspace $\cS_0$ corresponding to $\cN=0$ is spanned by four
vectors
\beq
\cS_0=\{\ket{000},\ket{001},\ket{010},\ket{011}\},
\label{cN0}
\eeq
where $\ket{0j_1 j_2}$ denotes a state with no photons in the
cavity and the atoms in state $\ket{j_1}_1\ket{j_2}_2$. The
restriction of $H_{\mathrm{meas}}$ to $\cS_0$ is the null operator
\beq
H_{\mathrm{meas}}|_{\cS_0} =0,
\eeq
hence $\cS_0$ is a subspace of the eigenspace $\cH_{P_0}$ of
$H_{\mathrm{meas}}$ belonging to the eigenvalue $\eta_0=0$
\beq
\cS_0 \subset \cH_{P_0}, \qquad
H_{\mathrm{meas}} P_0= 0 .
\eeq
The eigenspace $\cS_1$ corresponding to $\cN=1$ is spanned by
eight vectors
\beq
\cS_1=\{\ket{020},\ket{002},\ket{100},\ket{110},\ket{101},\ket{021},\ket{012},\ket{111}\},
\label{cN1}
\eeq
and the restriction of $H_{\mathrm{meas}}$ to $\cS_1$ is
represented by the 8-dimensional matrix
\beq
H_{\mathrm{meas}}|_{\cS_1} =\left(
\begin{array}{cccccccc}
  0 & 0 & 0 & ig & 0 & 0 & 0 & 0 \\
  0 & 0 & 0 & 0 & ig & 0 & 0 & 0 \\
  0 & 0 & -i\kappa & 0 & 0 & 0 & 0 & 0 \\
  -ig & 0 & 0 & -i\kappa & 0 & 0 & 0 & 0 \\
  0 & -ig & 0 & 0 & -i\kappa & 0 & 0 & 0 \\
  0 & 0 & 0 & 0 & 0 & 0 & 0 & ig \\
  0 & 0 & 0 & 0 & 0 & 0 & 0 & ig \\
  0 & 0 & 0 & 0 & 0 & -ig & -ig & -i\kappa
\end{array}\right) .
\eeq
The eigenvector $(\ket{021}-\ket{012})/\sqrt{2}$ has eigenvalue
$\eta_0=0$ and all the others have eigenvalues with negative
imaginary parts. Moreover, all restrictions
$H_{\mathrm{meas}}|_{\cS_n}$ with $n>1$ have eigenvalues with
negative imaginary parts. Indeed they are spanned by states
containing at least one photon, which dissipates through nonideal
mirrors, according to $-i\kappa b^\dagger b$ in (\ref{eq:cavity}).
The only exception is state $\ket{0,2,2}$ of $\cS_2$, but also in
this case it easy to prove that all eigenstates of
$H_{\mathrm{meas}}|_{\cS_2}$ dissipate. Therefore the eigenspace
$\cH_{P_0}$ of $H_{\mathrm{meas}}$ belonging to the eigenvalue
$\eta_0=0$ is 5-dimensional and is spanned by
\andy{decfree}
\beq
\cH_{P_0}=\{\ket{000},\ket{001},\ket{010},\ket{011},(\ket{021}-\ket{012})/\sqrt{2}\},
\label{decfree}
\eeq
 If the coupling $g$ and the cavity loss $\kappa$ are sufficiently strong,
any other weak Hamiltonian $H$ added to (\ref{eq:cavity}) reduces
to $P_0 H P_0$ and changes the state of the system only
\textit{within} the decoherence-free subspace (\ref{decfree}).

Fourth example. Let
\andy{hamqc}
\beq
H_{\mathrm{decay}}= H+ K H_{\mathrm{meas}}= \pmatrix{0 &
\tau_{\mathrm{Z}}^{-1} & 0\cr \tau_{\mathrm{Z}}^{-1} & -i 2/
\tau_{\mathrm{Z}}^2 \gamma & K \cr 0 & K & 0}.
\label{hamqc}
\eeq
This describes the spontaneous emission $\ket{1}\to\ket{2}$ of a
system into a (structured) continuum, while level $\ket{2}$ is
resonantly coupled to a third level $\ket{3}$.\cite{zenoreview}
This case is also relevant for quantum computation, if one is
interested in protecting a given subspace (level $\ket{1}$) from
decoherence\cite{Viola99,Beige00} by inhibiting spontaneous
emission.\cite{Agarwal01} Here $\gamma$ represents the decay rate
to the continuum and $\tau_{\mathrm{Z}}$ is the Zeno time
(convexity of the initial quadratic region).

Notice that, in a certain sense, this situation is complementary
to that in (\ref{eq:cavity}); here the measurement Hamiltonian
$H_{\mathrm{meas}}$ is Hermitian, while the system Hamiltonian $H$
is not. Again, we have to enlarge our Hilbert space, apply the
theorem to the dilation and project back the Zeno evolution. As a
result one can simply apply the theorem to the original
Hamiltonian, for, in this case, $H_{\mathrm{meas}}$ has a complete
set of orthogonal projections that univocally defines a partition
of $\cH$ into quantum Zeno subspaces. We shall elaborate further
on this interesting aspect in a future work.

 As the Rabi frequency $K$ is
increased one is able to hinder spontaneous emission from level
$\ket{1}$ (to be protected) to level $\ket{2}$. However, in order
to get an effective ``protection" of level $\ket{1}$, one needs $K
> 1/\tau_{\mathrm{Z}}$. More to this, when the presence of the
inverse Zeno effect is taken into account, this requirement
becomes even more stringent\cite{heraclitus} and yields $K >
1/\tau_{\mathrm{Z}}^2\gamma$. Both these conditions can be very
demanding for a real system subject to
dissipation.\cite{zenoreview,Napoli,heraclitus} For instance,
typical values for spontaneous decay in vacuum are $\gamma\simeq
10^9$s$^{-1}$, $\tau_{\mathrm{Z}}^2\simeq 10^{-29}$s$^2$ and
$1/\tau_{\mathrm{Z}}^2\gamma\simeq
10^{20}$s$^{-1}$.\cite{hydrogen}

\section{Conclusions}
\label{sec-concl}

If very frequent measurements are performed on a quantum system,
in order to ascertain whether it is still in its initial state,
transitions to other states are hindered and the QZE takes place.
This formulation of the QZE hinges upon the notion of pulsed
measurements, according to von Neumann's projection postulate.
However, as we have seen by means of several examples, a
``measurement" is nothing but an interaction with an external
system (another quantum object, or a field, or simply another
degree of freedom of the very system investigated), playing the
role of apparatus. This enables one to reformulate the QZE in
terms of a (strong) coupling to an external agent and to cast the
quantum Zeno evolution in terms of an adiabatic theorem. There are
many interesting examples, varying from quantum computation to
decoherence-free subspaces to ``protection" from decoherence.
Additional work is in progress, also in view of possible practical
applications.

\section*{Acknowledgments}
I am grateful to Saverio Pascazio for helpful discussions.

\end{document}